\documentclass[final,5p,times,twocolumn]{elsarticle}
\usepackage{amsmath,amssymb,mathtools}
\usepackage{bm}
\usepackage{graphicx}
\usepackage{booktabs}
\usepackage{siunitx}
\usepackage{microtype}
\usepackage{hyperref}
\usepackage{xcolor}

\ifdefined\CSFSubmission
  \usepackage{lineno}
\fi

\journal{Chaos, Solitons \& Fractals}

\newcommand{\PT}{\mathcal{PT}}
\newcommand{\ii}{\mathrm{i}}
\newcommand{\dd}{\mathrm{d}}
\newcommand{\khat}{\widehat{\kappa}}
\newcommand{\safeimage}[2]{%
  \IfFileExists{#1}{\includegraphics[width=#2]{#1}}{%
  \fbox{\parbox[c][4.0cm][c]{#2}{\centering\footnotesize Figure file not supplied}}}}

\begin{document}

\begin{frontmatter}

\title{Re-entrant parity--time phase transitions in locally coupled ring resonators}

\author[HUST]{Nguyen Duc Anh Quan}
\author[VLU,VLU1]{Le Xuan The Tai}
\author[HUST]{Doan Quang Tri}
\author[UM,PW]{Pawel S. Jung}
\author[UW]{Marek Trippenbach}
\author[HUST]{Nguyen Viet Hung\corref{cor1}}

\cortext[cor1]{Corresponding author.}

\ead{hung.nguyenviet1@hust.edu.vn}

\affiliation[HUST]{
  organization={School of Materials Science and Engineering, Hanoi University of Science and Technology},
  addressline={1 Dai Co Viet Street},
  city={Hanoi},
  postcode={100000},
  country={Vietnam}
}
\affiliation[VLU]{
  organization={Atomic, Molecular and Optical Physics Research Group, Science and Technology Advanced Institute, Van Lang University},
  addressline={69/68 Dang Thuy Tram},
  city={Ho Chi Minh City},
  postcode={70000},
  country={Vietnam}
}
\affiliation[VLU1]{
  organization={Faculty of Applied Technology, Van Lang School of Technology, Van Lang University},
  addressline={69/68 Dang Thuy Tram},
  city={Ho Chi Minh City},
  postcode={70000},
  country={Vietnam}
}
\affiliation[UM]{
  organization={Department of Physics},
  addressline={University of Miami},
  city={Coral Gables},
  state={FL},
  postcode={33146},
  country={USA}
}
\affiliation[PW]{
  organization={Faculty of Physics, Warsaw University of Technology},
  addressline={Koszykowa 75},
  city={Warsaw},
  postcode={00-662},
  country={Poland}
}
\affiliation[UW]{
  organization={Faculty of Physics, University of Warsaw},
  addressline={Pasteura 5},
  city={Warsaw},
  postcode={02-093},
  country={Poland}
}

\begin{abstract}
We investigate two parity--time-symmetric ring resonators coupled over a
finite angular region described by a super-Gaussian profile. In the linear
regime, analytical spectra are obtained in the homogeneous-coupling and
fixed-amplitude narrow-contact limits, while the finite-width problem is
treated numerically. Local coupling introduces nonzero spatial Fourier
components that mix angular harmonics and lift the degeneracy of
counterpropagating modes, resolving each excited doublet into
parity-dependent branches. Collisions among these branches generate multiple
exceptional-point boundaries and disconnected broken-$\PT$ domains. The
resulting phase diagrams exhibit re-entrant
unbroken--broken--unbroken transitions when the gain--loss strength,
coupling width, or peak coupling amplitude is varied. The numerical spectra
continuously recover both analytical limits. In the nonlinear regime,
selected ground and excited linear modes are used as seeds for adiabatic
propagation into finite-amplitude Kerr waveforms that remain dynamically
persistent over the simulated observation interval for finite ranges of
nonlinear strength. These results show that the spatial profile of
inter-resonator coupling provides a geometric means of controlling
multimode $\PT$ transitions and selecting dynamically accessible nonlinear
waveforms in coupled-ring systems.
\end{abstract}


\end{frontmatter}

\ifdefined\CSFSubmission
  \linenumbers
\fi

\section{Introduction}
\label{sec:introduction}

Non-Hermitian systems with parity--time ($\PT$) symmetry provide a framework
in which a non-Hermitian operator can possess an entirely real spectrum
within a finite parameter domain
\cite{Bender1998,Bender1999,Bender2007}. The boundaries of this domain are
formed by exceptional points, where both eigenvalues and their associated
eigenvectors coalesce \cite{Heiss2012,Miri2019}. Optical systems are
particularly suitable for investigating these phenomena because
refractive-index landscapes, gain, loss, and coupling can be engineered with
considerable flexibility. Early coupled-mode and lattice models established
the basic optical consequences of balanced gain and loss
\cite{ElGanainy2007,Musslimani2008,Makris2008}, followed by experimental
demonstrations of $\PT$-symmetry breaking in optical waveguides
\cite{Guo2009,Ruter2010}. Nonlinearity further enriches the dynamics by
enabling asymmetric stationary states, directional transport, solitons,
breathers, and nonlinear symmetry transitions
\cite{Ramezani2010,Miroshnichenko2011,Alexeeva2012,Nixon2012,
Barashenkov2012,Lumer2013}. Broader accounts of nonlinear and
non-Hermitian photonics can be found in
Refs.~\cite{Konotop2016,Suchkov2016,ElGanainy2018,Ozdemir2019}.

Optical microresonators provide compact, high-quality-factor platforms for
controlling modal spectra, coupling, dispersion, and nonlinear response
\cite{Vahala2003,Kippenberg2018}. Coupled microresonators are therefore
natural systems for studying the competition among inter-resonator coupling,
gain, and loss. $\PT$ transitions and exceptional-point phenomena have been
demonstrated in whispering-gallery resonators, active--passive microcavity
pairs, and microring lasers
\cite{Peng2014NP,Chang2014,Hodaei2014}. Nonuniform pumping provides an
additional route to pump-induced exceptional points in coupled laser
cavities \cite{Liertzer2012}. Related experiments have demonstrated
loss-induced suppression and revival of lasing \cite{Peng2014Science},
mode selection through $\PT$-symmetry breaking
\cite{Feng2014,Hodaei2014}, and reversal of the pump dependence near an
exceptional point \cite{Brandstetter2014}. Exceptional points have also
attracted interest for their enhanced spectral response to weak
perturbations \cite{Chen2017EP} and for asymmetric mode conversion through
dynamical encircling \cite{Doppler2016EP}. These developments motivate
methods for controlling the location and multiplicity of exceptional points
in multimode resonator systems.

Most elementary descriptions of a $\PT$-symmetric dimer assume a single
spatially uniform coupling coefficient. This approximation is appropriate
when the evanescent interaction extends over the entire circumference or
when the longitudinal structure is projected onto a single mode. In
realistic ring geometries, however, two resonators may interact only over a
finite angular sector. The resulting local coupling contains nonzero spatial
Fourier components that mix different angular harmonics. It also removes the
continuous rotational symmetry responsible for the degeneracy of clockwise
and counterclockwise modes, while an even coupling profile preserves
reflection symmetry. More generally, degeneracy can qualitatively modify
multimode $\PT$ transitions and may permit the restoration of real spectral
sectors, whereas cyclic waveguide networks provide a related setting in
which coupling geometry controls symmetry breaking
\cite{GeStone2014,Barashenkov2013Necklace}. Previous studies of coupled ring
systems have demonstrated modulational instability, vortical states,
symmetry breaking, and routes to chaos in the presence of linear gain and
nonlinear loss \cite{Hung2017,Ramaniuk2018,Zegadlo2019,Nguyen2019}. These
works show that the spatial structure of the coupling region can strongly
influence the dynamics. Nevertheless, a systematic characterization of how
a continuously tunable local coupling profile lifts the
counterpropagating-mode degeneracy, creates multiple exceptional-point
boundaries, and enables re-entrant $\PT$ transitions in a balanced ring pair
is still lacking.

Here, we address this problem by considering two identical rings carrying
equal-magnitude gain and loss and coupled through an even, high-order
super-Gaussian profile. In the homogeneous-coupling and fixed-amplitude
narrow-contact limits, the model admits simple analytical spectra that
provide stringent checks on the finite-width numerical calculations. For
intermediate coupling widths, the nonzero Fourier components of the local
interaction mix angular harmonics and lift the degeneracy between clockwise
and counterclockwise modes, resolving each excited doublet into four
parity-dependent branches. Collisions among these branches generate
multiple exceptional-point boundaries and disconnected broken-$\PT$
domains. The phase transition is therefore not governed by a single
monotonic threshold: within suitable parameter intervals, variation of the
gain--loss strength, coupling width, or peak coupling amplitude produces an
unbroken--broken--unbroken sequence. We further use selected ground and
excited linear modes as seeds for adiabatic nonlinear propagation. The
resulting finite-amplitude Kerr waveforms remain dynamically persistent over
the simulated observation interval for finite ranges of nonlinear strength,
with a substantially narrower accessible range for the excited branch.

The paper is organized as follows. Section~\ref{sec:model} introduces the
model, its $\PT$ symmetry, and the numerical diagnostics.
Section~\ref{sec:linear} derives the two analytical limits and analyzes the
finite-width spectrum, phase maps, and re-entrant transitions.
Section~\ref{sec:nonlinear} presents the nonlinear continuation procedure
and the Kerr results, including a brief comparison with a saturable
response. Section~\ref{sec:conclusion} summarizes the principal conclusions
and clarifies the scope of the nonlinear stability statements.
\section{Model and numerical formulation}
\label{sec:model}

\subsection{Evolution equations and symmetry}

We consider two identical ring resonators parametrized by the dimensionless angular coordinate $x\in[-\pi,\pi)$ and subject to periodic boundary conditions. The slowly varying fields $\psi_1(x,t)$ and $\psi_2(x,t)$ obey
\begin{align}
\ii\frac{\partial \psi_1}{\partial t}
&=-\frac{\partial^2\psi_1}{\partial x^2}
+\ii\gamma\psi_1
+\frac{\sigma |\psi_1|^2}{1+\beta |\psi_1|^2}\psi_1
+\kappa(x)\psi_2,
\label{eq:model1}\\
\ii\frac{\partial \psi_2}{\partial t}
&=-\frac{\partial^2\psi_2}{\partial x^2}
-\ii\gamma\psi_2
+\frac{\sigma |\psi_2|^2}{1+\beta |\psi_2|^2}\psi_2
+\kappa(x)\psi_1.
\label{eq:model2}
\end{align}
The diffraction/dispersion coefficient and ring radius have been scaled to unity. The gain--loss coefficient is $\gamma\geq0$, with gain in ring 1 and the same amount of loss in ring 2. The real coefficient $\sigma$ controls the nonlinear phase shift. The choice $\beta=0$ gives the Kerr response studied in detail. The case $\beta=1$ is used only for a brief supporting comparison with a saturable response; both are standard local nonlinear-optical models \cite{Boyd2020,Yang2010}. The coupling profile is
\begin{equation}
\kappa(x)=\kappa_0\exp\!\left[-\left(\frac{x}{W}\right)^{2m}\right],
\qquad m=10,
\label{eq:coupling}
\end{equation}
where $\kappa_0$ is the peak coupling and $W$ sets the angular extent of the coupling region. The large value of $m$ gives a smooth approximation to a coupling window that is almost flat and has steep edges. We also verified numerically that the qualitative spectral behavior persists
for other smooth localized coupling profiles, including a Gaussian profile.

\begin{figure}[!t]
\centering
\safeimage{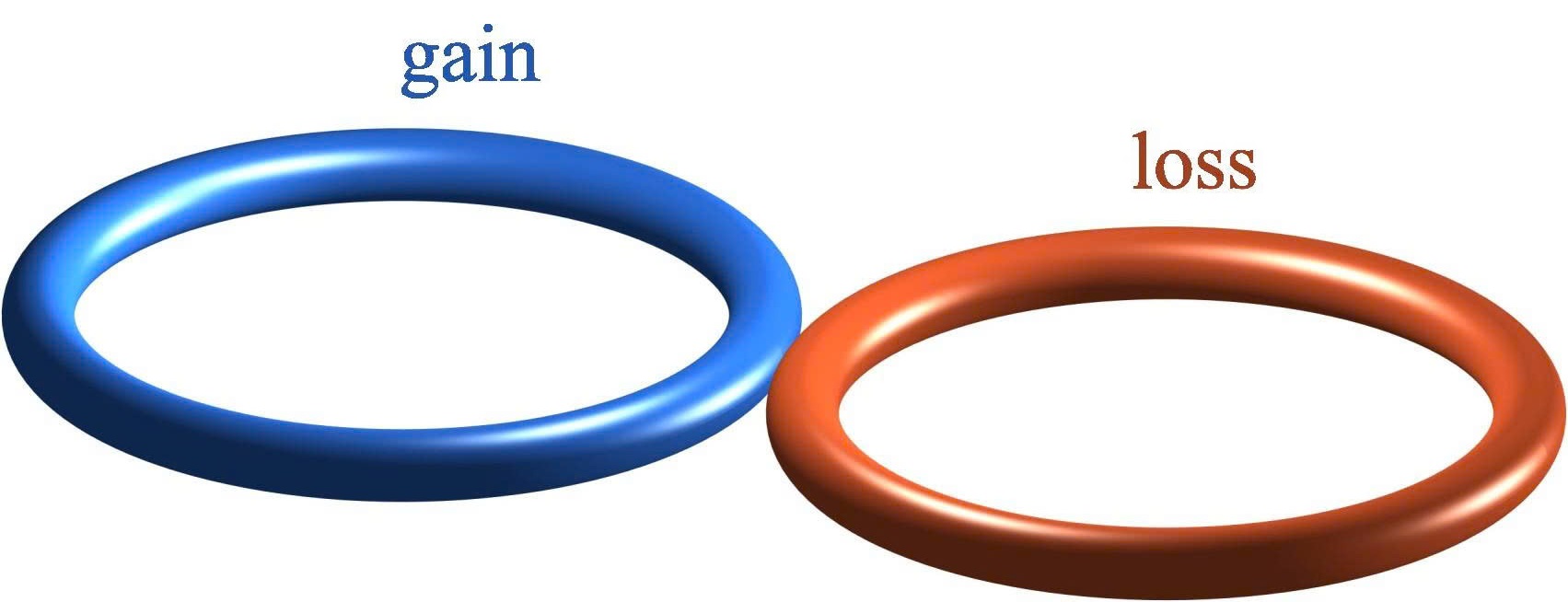}{0.96\linewidth}
\caption{Schematic representation of two locally coupled ring resonators. Ring 1 carries linear gain $+\gamma$, ring 2 carries the balanced linear loss $-\gamma$, and the real inter-ring coupling is concentrated in a finite angular sector described by Eq.~\eqref{eq:coupling}.}
\label{fig:scheme}
\end{figure}

Because $\kappa(x)$ is real and even, Eqs.~\eqref{eq:model1}--\eqref{eq:model2} are invariant under the combined operation
\begin{equation}
\begin{aligned}
\mathcal{P}:\quad &\{\psi_1(x),\psi_2(x)\}\mapsto\{\psi_2(-x),\psi_1(-x)\},\\
\mathcal{T}:\quad &\ii\mapsto-\ii,\qquad t\mapsto-t.
\end{aligned}
\notag 
\end{equation}
Thus, in the linear problem the spectrum is either real or arranged in complex-conjugate pairs. In the nonlinear problem, a $\PT$-symmetric stationary state satisfies $\phi_2(x)=\phi_1^*(-x)$ up to a constant phase convention.

The powers in the individual rings and the total power are
\begin{equation}
P_j(t)=\int_{-\pi}^{\pi}|\psi_j(x,t)|^2\,\dd x,
\qquad P(t)=P_1(t)+P_2(t).
\label{eq:powers}
\end{equation}
Directly from Eqs.~\eqref{eq:model1}--\eqref{eq:model2},
\begin{equation}
\frac{\dd P}{\dd t}=2\gamma\,[P_1(t)-P_2(t)].
\label{eq:powerbalance}
\end{equation}
The total power is therefore constant for a perfectly balanced $\PT$-symmetric stationary state, although it may vary during an adiabatic nonlinear ramp while the field readjusts.

\subsection{Linear eigenproblem and spectral diagnostic}

Setting $\sigma=0$ and using $\bm{\psi}(x,t)=\bm{\phi}(x)e^{-\ii\lambda t}$ gives
\begin{equation}
\mathcal{H}\bm{\phi}=\lambda\bm{\phi},
\qquad
\mathcal{H}=
\begin{pmatrix}
-\partial_x^2+\ii\gamma & \kappa(x)\\
\kappa(x) & -\partial_x^2-\ii\gamma
\end{pmatrix}.
\label{eq:linearH}
\end{equation}
For finite $W$, we discretize the periodic domain by Fourier collocation \cite{Trefethen2000} and diagonalize the resulting non-Hermitian matrix. Equivalently, expanding
\begin{equation}
\begin{aligned}
\phi_1(x)&=\sum_n a_n e^{\ii n x},\qquad
\phi_2(x)=\sum_n b_n e^{\ii n x},\\
\kappa(x)&=\sum_q \khat_q e^{\ii qx}.
\end{aligned}
\label{eq:fourierexpansion}
\end{equation}
with
\begin{equation}
\khat_q=\frac{1}{2\pi}\int_{-\pi}^{\pi}\kappa(x)e^{-\ii qx}\,\dd x,
\label{eq:kappahat}
\end{equation}
leads to
\begin{align}
\lambda a_n&=(n^2+\ii\gamma)a_n+\sum_l\khat_{n-l}b_l,
\label{eq:fourier1}\\
\lambda b_n&=(n^2-\ii\gamma)b_n+\sum_l\khat_{n-l}a_l.
\label{eq:fourier2}
\end{align}
These equations make the physical role of local coupling transparent: a uniform coupler has only $\khat_0$, whereas a local coupler contains nonzero harmonics $\khat_{q\neq0}$ and mixes different angular momenta.

We characterize the linear phase by the spectral growth rate
\begin{equation}
\Gamma_{\mathrm{max}}(W,\kappa_0,\gamma)
=\max_j \{\operatorname{Im} (\lambda_j)\}.
\label{eq:gammaMax}
\end{equation}
The system is in the unbroken-$\PT$ phase when all eigenvalues are real within numerical tolerance, and in the broken phase when $\Gamma_{\mathrm{max}}>0$. Since complex eigenvalues occur in conjugate pairs, the largest positive imaginary part is sufficient to map the broken regions.

\section{Linear locally coupled rings}
\label{sec:linear}

\subsection{Homogeneous-coupling limit}
\label{sec:homogeneous}

When $W$ is much larger than the ring circumference, Eq.~\eqref{eq:coupling} becomes effectively constant, $\kappa(x)=\kappa_0$. A plane wave
\begin{equation}
\psi_j(x,t)=A_j e^{\ii n x-\ii\lambda t},\qquad n\in\mathbb{Z},
\label{eq:planeWave}
\end{equation}
reduces the eigenproblem to
\begin{equation}
\begin{pmatrix}
n^2+\ii\gamma & \kappa_0\\
\kappa_0 & n^2-\ii\gamma
\end{pmatrix}
\begin{pmatrix}A_1\\A_2\end{pmatrix}
=\lambda\begin{pmatrix}A_1\\A_2\end{pmatrix}.
\label{eq:uniformMatrix}
\end{equation}
The characteristic equation is
\begin{equation}
(\lambda-n^2)^2=\kappa_0^2-\gamma^2,
\end{equation}
and hence
\begin{equation}
\lambda_{n,\pm}=n^2\pm\sqrt{\kappa_0^2-\gamma^2}.
\label{eq:uniformSpectrum}
\end{equation}
The homogeneous system has an entirely real spectrum for
\begin{equation}
\gamma\leq\kappa_0,
\label{eq:uniformThreshold}
\end{equation}
with an exceptional point at $\gamma=\kappa_0$. For $\gamma=0$, the two ring supermodes are shifted by $\pm\kappa_0$ relative to the single-ring rotor spectrum $n^2$. Because Eq.~\eqref{eq:uniformSpectrum} depends on $n^2$, the clockwise and counterclockwise states $\pm n$ are degenerate for every $n\geq1$.

\subsection{Vanishing-width, fixed-amplitude contact}
\label{sec:narrow}

At fixed peak coupling $\kappa_0$, the integrated strength of the super-Gaussian profile vanishes linearly with $W$. Indeed, for $W\ll\pi$,
\begin{equation}
\int_{-\pi}^{\pi}\kappa(x)\,\dd x
\simeq \frac{\kappa_0 W}{m}\,
\Gamma\!\left(\frac{1}{2m}\right)
\longrightarrow0.
\label{eq:couplingArea}
\end{equation}
Thus, the $W\to0$ limit of Eq.~\eqref{eq:coupling} is a vanishing-area point contact and the two rings become asymptotically independent. Their spectra are
\begin{equation}
\lambda_{n,\pm}=n^2\pm\ii\gamma.
\label{eq:narrowSpectrum}
\end{equation}
For any $\gamma>0$, one branch grows and its conjugate decays, so the linear system is in the broken-$\PT$ phase. The $\pm n$ degeneracy remains because each isolated ring still has rotational symmetry.

It should be emphasized that Eq.~\eqref{eq:narrowSpectrum} corresponds to the fixed-amplitude profile used throughout this work. A mathematically normalized Dirac delta interaction would require $\kappa_0\propto W^{-1}$ so that the area under $\kappa(x)$ remains finite  as $W \to 0$. That is a different limiting problem, with derivative jump conditions at the contact, and is not the limit followed by the numerical data presented here.

\subsection{Finite-width coupling and lifting of angular degeneracy}
\label{sec:finiteWidth}

For finite $W$, the local coupling has nonzero Fourier harmonics and angular momentum is no longer conserved separately. The residual reflection symmetry $x\mapsto-x$ nevertheless allows the modes to be classified by even and odd spatial parity. Within the degenerate subspace spanned by $e^{\ii n x}$ and $e^{-\ii n x}$, the component $\khat_{2n}$ directly couples the counterpropagating waves. Their even and odd combinations, proportional to $\cos(nx)$ and $\sin(nx)$, therefore acquire different shifts. Combining this parity splitting with the two inter-ring supermode branches gives four branches for every excited doublet $n\geq1$. The ground state $n=0$ has no counterpropagating partner and retains only two branches. The complete finite-width spectrum also contains mixing with neighboring angular harmonics through the remaining coefficients $\khat_q$, so the branch positions are determined by the full local-coupling matrix rather than by $\khat_{2n}$ alone.

\begin{figure*}[!t]
\centering
\safeimage{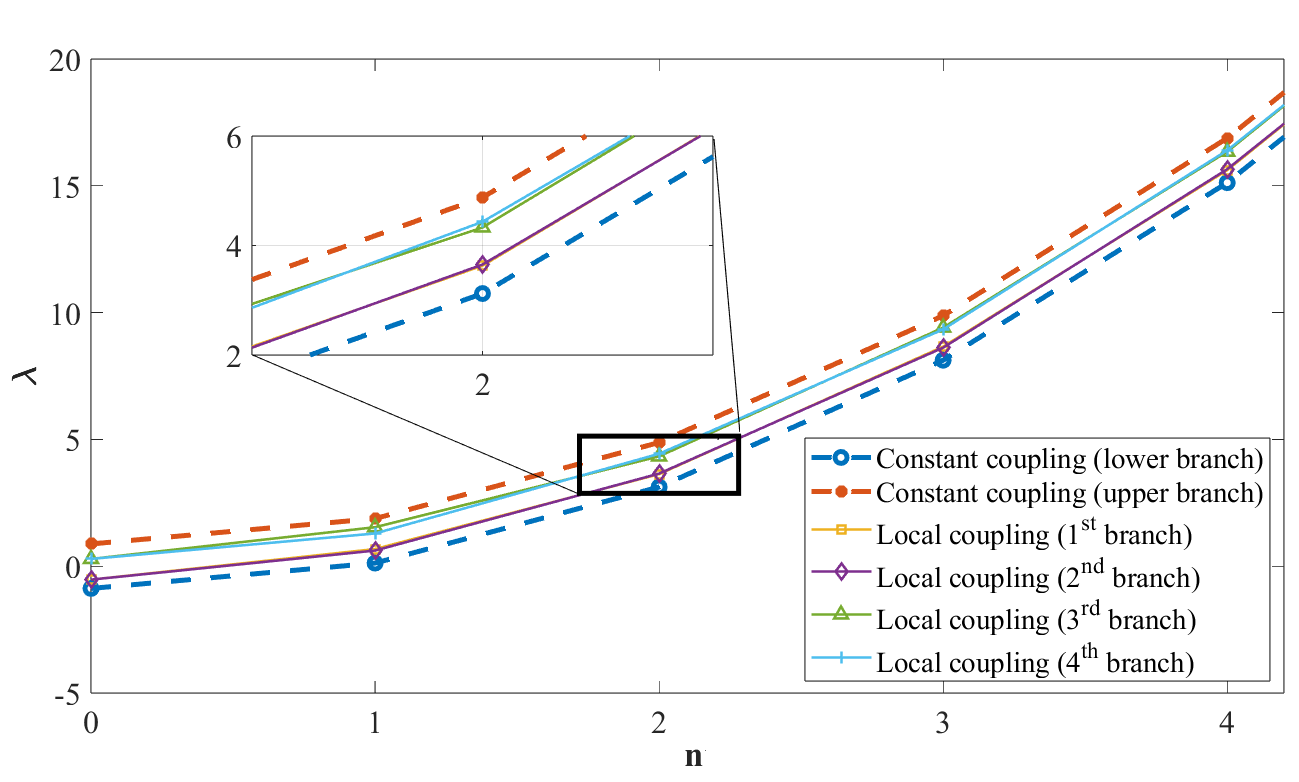}{0.82\textwidth}
\caption{Hermitian spectrum at $\kappa_0=0.9$ and $\gamma=0$. The two dashed outer branches are the homogeneous-coupling values, each retaining the $\pm n$ degeneracy. For the local profile with $W=1.8$, each excited doublet is resolved into four parity-dependent branches. Broadening the coupling region suppresses the nonzero Fourier harmonics and recombines these branches into the homogeneous spectrum.}
\label{fig:HermitianSpectrum}
\end{figure*}

Figure~\ref{fig:HermitianSpectrum} compares the Hermitian local-coupling spectrum with the homogeneous result. The two outer branches are the doubly degenerate values from Eq.~\eqref{eq:uniformSpectrum}, while the four inner local-coupling branches resolve the even and odd members of each $\pm n$ doublet. All eigenvalues remain real because $\gamma=0$. As the coupling window broadens, the nonzero Fourier components of $\kappa(x)$ are suppressed and the four local branches continuously recombine into the two homogeneous branches.

\begin{figure*}[!t]
\centering
\safeimage{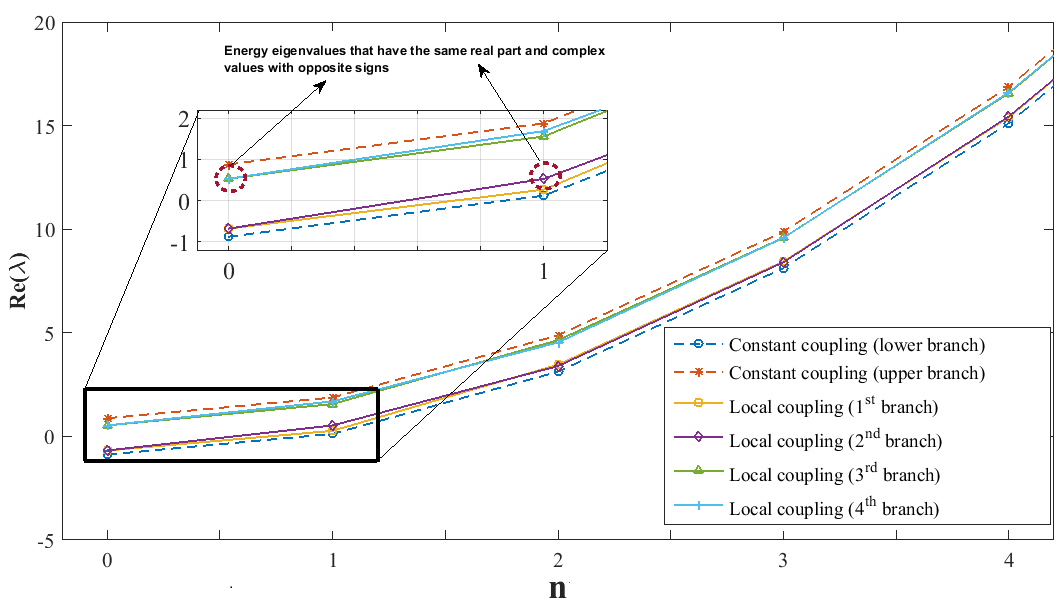}{0.82\textwidth}
\caption{Real parts of the non-Hermitian spectrum at $\kappa_0=0.9$ and $\gamma=0.2$, comparing the local profile with $W=2.2$ and the homogeneous limit. Local coupling again resolves each excited doublet into four branches. The annotated pairs have coincident real parts and nonzero imaginary parts of opposite sign in the full numerical spectrum; they therefore represent broken-$\PT$ complex-conjugate eigenvalues.}
\label{fig:NonHermitianSpectrum}
\end{figure*}

The same four-branch organization remains visible in the non-Hermitian spectrum of Fig.~\ref{fig:NonHermitianSpectrum}. Here the vertical axis displays $\operatorname{Re}\lambda$. The annotations identify branch pairs that have the same real part while the corresponding numerical eigenvalues possess imaginary parts of equal magnitude and opposite sign. Thus, the appearance of a common real part is accompanied by a complex-conjugate pair in the full spectrum. This distinction is important: Fig.~\ref{fig:NonHermitianSpectrum} visualizes the rearrangement of the real parts, whereas the broken-$\PT$ classification is made from the complete complex eigenvalues through Eq.~\eqref{eq:gammaMax}.

\begin{figure}[!t]
\centering
\safeimage{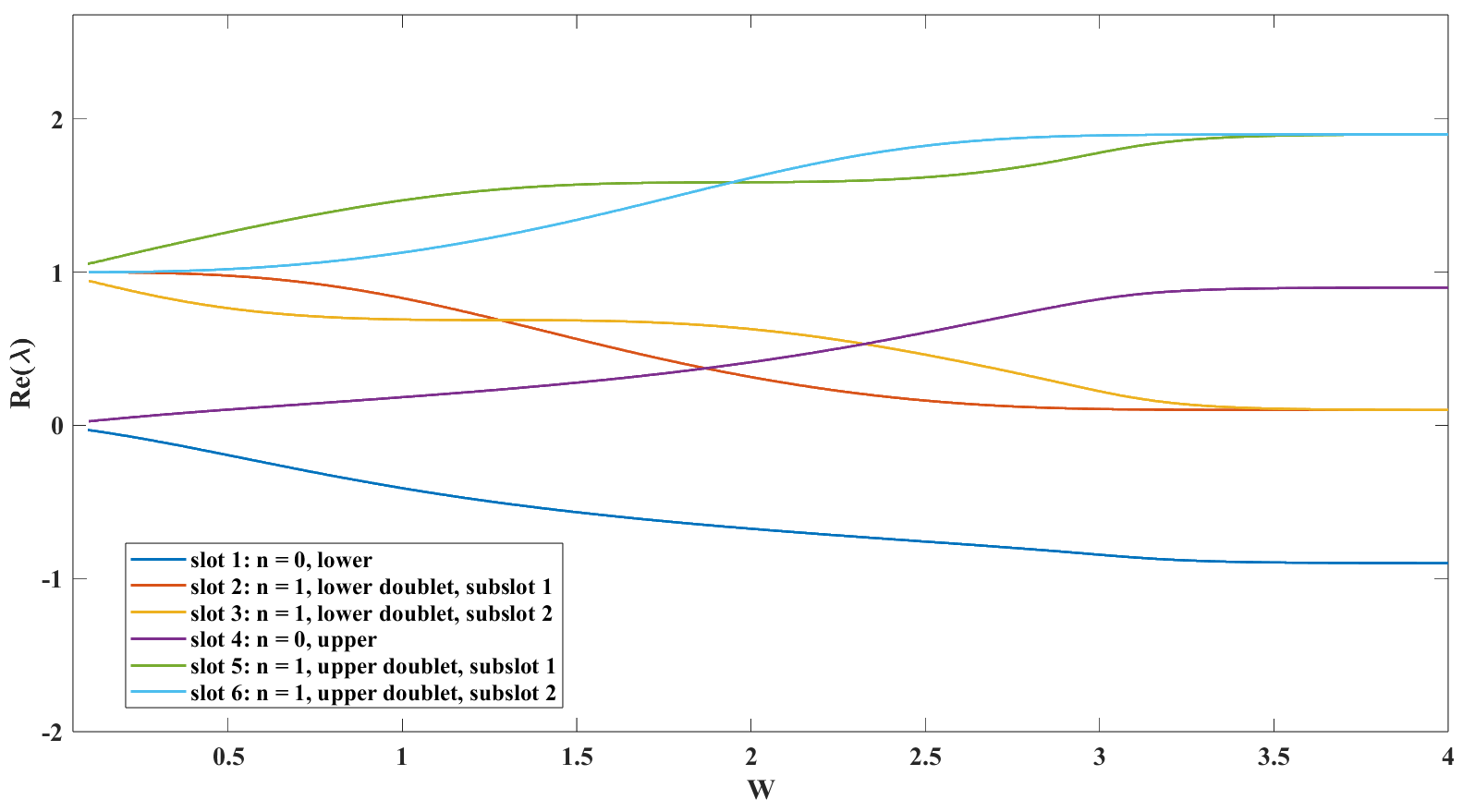}{0.98\columnwidth}
\caption{Six lowest Hermitian eigenvalues versus coupling width at $\kappa_0=0.9$ and $\gamma=0$. All branches remain real; symmetry-distinct states cross without coalescence. The large- and small-$W$ limits approach Eqs.~\eqref{eq:uniformSpectrum} and \eqref{eq:narrowSpectrum}, respectively.}
\label{fig:HermitianWidth}
\end{figure}

A convenient way to follow the spectral rearrangement is to start at large $W$, where each branch can be assigned to an analytical value in Eq.~\eqref{eq:uniformSpectrum}, and then decrease $W$ continuously. Figure~\ref{fig:HermitianWidth} shows that the six lowest Hermitian branches remain real and can cross without forming defective states. For large values of $W$, the two homogeneous supermode levels are recovered, whereas for small values of $W$, the spectrum approaches the degenerate isolated-ring rotor spectrum.

\begin{figure}[!t]
\centering
\safeimage{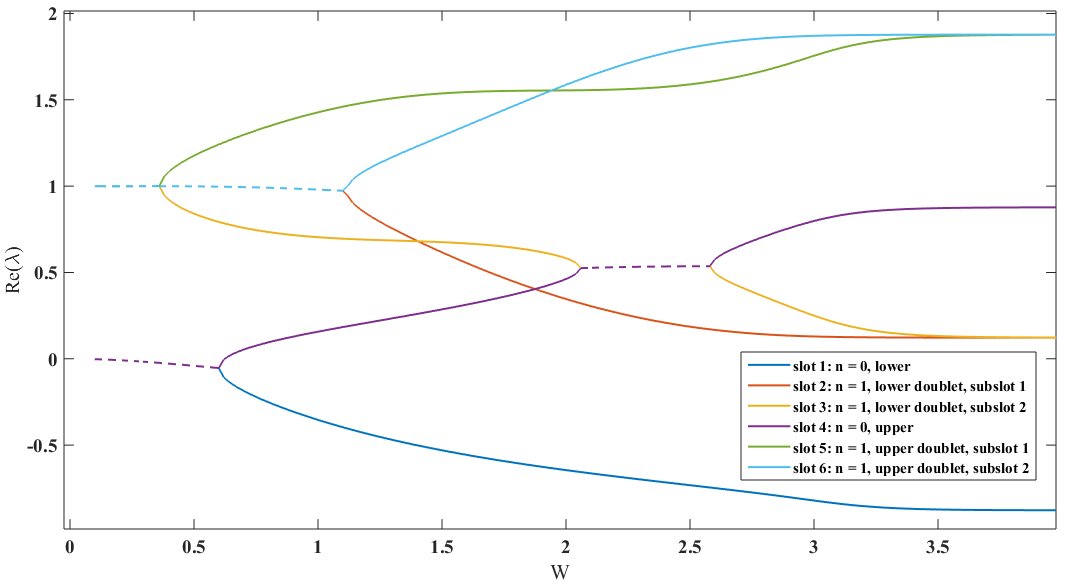}{0.98\columnwidth}
\caption{Real parts of the six lowest non-Hermitian eigenvalues versus coupling width at $\kappa_0=0.9$ and $\gamma=0.2$. Dashed segments mark complex-conjugate pairs with equal real parts and opposite imaginary parts. Their endpoints are exceptional points. A detailed view of the evolution of the eigenvalues and the corresponding eigenstates is provided in the \textbf{\textit{Supplementary Movies S1--S2}}.}
\label{fig:NonHermitianWidth}
\end{figure}

The non-Hermitian evolution in Fig.~\ref{fig:NonHermitianWidth} is qualitatively different. The plotted quantity is again $\operatorname{Re}\lambda$. We added dashes to highlight the new features. Over each dashed interval, two branches share the same real part while the full eigenvalues form a conjugate pair with imaginary parts $\pm\operatorname{Im}\lambda$. At either end of such an interval, the two eigenvalues and their eigenvectors coalesce at an exceptional point; outside it they separate onto the real axis. When $W$ is decreased further, the numerical spectrum approaches Eq.~\eqref{eq:narrowSpectrum}: $\operatorname{Re}\lambda\to n^2$ while $\operatorname{Im}\lambda\to\pm\gamma$. Recovering both limits in the same branch-tracking calculation provides a direct consistency check on the discretization and eigenvalue assignment. The evolution of the eigenvalues and the corresponding eigenstates as W varies is illustrated in the \textbf{\textit{Supplementary Movies $S_1$ - $S_2$}}.

\subsection{Two-parameter phase maps}
\label{sec:phaseMaps}

\begin{figure*}[!t]
\centering
\safeimage{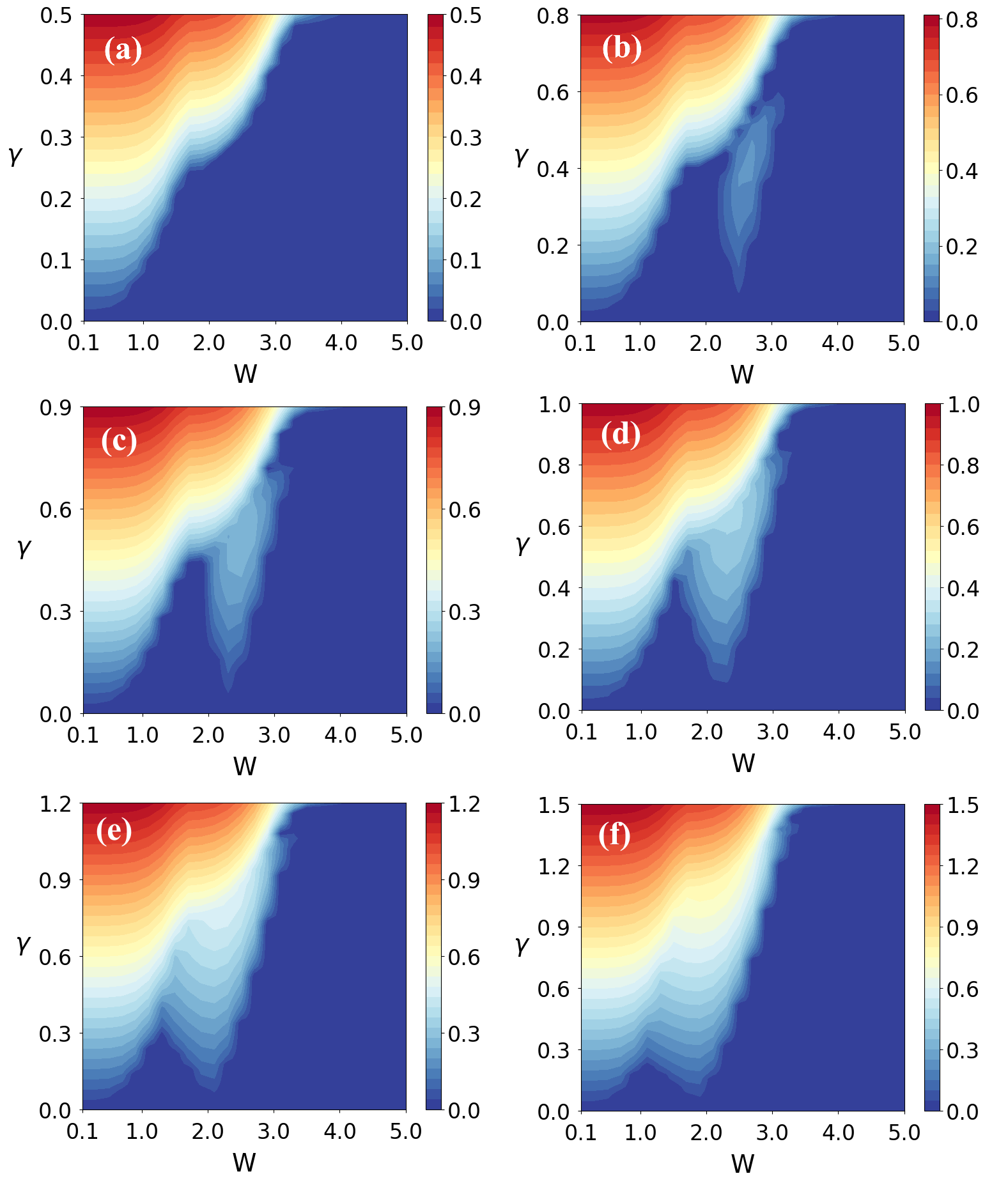}{0.74\textwidth}
\caption{Maximum spectral growth rate $\Gamma_{\mathrm{max}}=\max_j \{\operatorname{Im} (\lambda_j)\}$ in the $(W,\gamma)$ plane for (a) $\kappa_0=0.5$, (b) $0.8$, (c) $0.9$, (d) $1.0$, (e) $1.2$, and (f) $1.5$. Dark-blue regions have $\Gamma_{\mathrm{max}}=0$ and are unbroken-$\PT$; positive values mark broken-$\PT$ domains. Each panel uses the color range shown on its own colorbar, so cross-panel comparison should focus on the boundary geometry. The maps recover the broken narrow-contact limit, approach $\gamma=\kappa_0$ at large $W$, and reveal intermediate tongues and islands generated by distinct modal coalescences.}
\label{fig:phaseMaps}
\end{figure*}

To obtain a global view, we diagonalize Eq.~\eqref{eq:linearH} over the $(W,\gamma)$ plane and plot $\Gamma_{\mathrm{max}}$ for several values of $\kappa_0$. In Fig.~\ref{fig:phaseMaps}, dark blue denotes $\Gamma_{\mathrm{max}}=0$ within the numerical tolerance, while lighter and warmer colors indicate positive growth rates. The color range of each panel follows its displayed $\kappa_0$ value; consequently, the geometry of the phase boundaries can be compared directly across panels, whereas equal colors in different panels do not necessarily represent equal absolute growth rates.

The two analytical limits are recovered throughout the sequence. Near the narrow-contact edge, any nonzero $\gamma$ produces a broken spectrum, as predicted by Eq.~\eqref{eq:narrowSpectrum}. On the large-$W$ side, the numerical boundary tends toward the homogeneous threshold $\gamma=\kappa_0$. The finite-width region between them is not a smooth interpolation. For $\kappa_0=0.5$ the boundary is comparatively simple, but at $\kappa_0=0.8$ a narrow broken-$\PT$ tongue appears inside the otherwise real-spectrum domain around intermediate $W$. This feature expands and deforms for $\kappa_0=0.9$ and $1.0$, producing pronounced re-entrant cuts through the phase diagram. At still larger coupling, the separate structures merge into a broader lobe, although the boundary retains a strong indentation relative to the homogeneous result.

The lobes and islands are the global traces of different modal collisions. Because the Fourier components of $\kappa(x)$ shift the even and odd sectors by different amounts and mix neighboring angular harmonics, distinct branch pairs reach their exceptional points along different curves in the $(W,\gamma)$ plane. Their superposition replaces the single dimer threshold by several exceptional-point contours. This behavior is geometric rather than topological--no winding number or topological invariant is introduced--but it establishes the coupling width as an independent spectral control parameter capable of selecting different $\PT$-phase sequences at fixed gain and loss.

\subsection{Re-entrant $\PT$ transitions}
\label{sec:reentrant}

The homogeneous dimer has a single threshold, Eq.~\eqref{eq:uniformThreshold}. Local coupling replaces that threshold by a multimode set of exceptional-point boundaries. The reason is that the Fourier harmonics of $\kappa(x)$ shift the even and odd sectors differently and couple neighboring angular modes. As a control parameter is varied, one pair of branches may coalesce, become complex, and then separate again before another pair reaches an exceptional point. The result is a local re-entrant sequence rather than a single irreversible transition.

\begin{figure*}[!t]
\centering
\safeimage{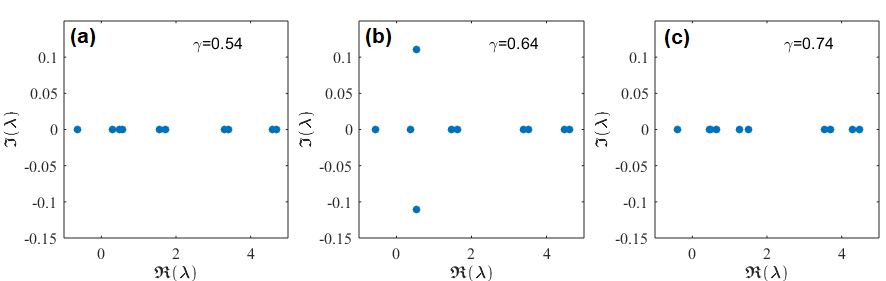}{0.9\linewidth}\hfill
\safeimage{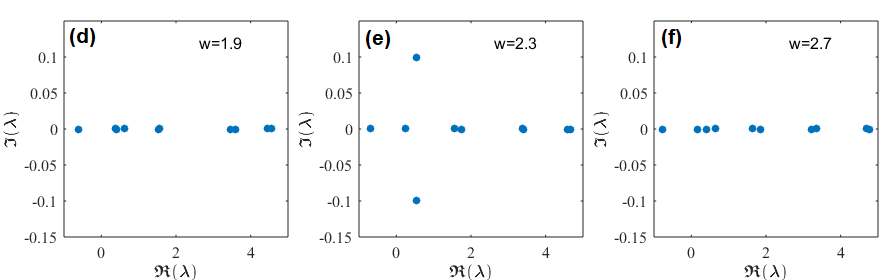}{0.9\linewidth}\hfill
\safeimage{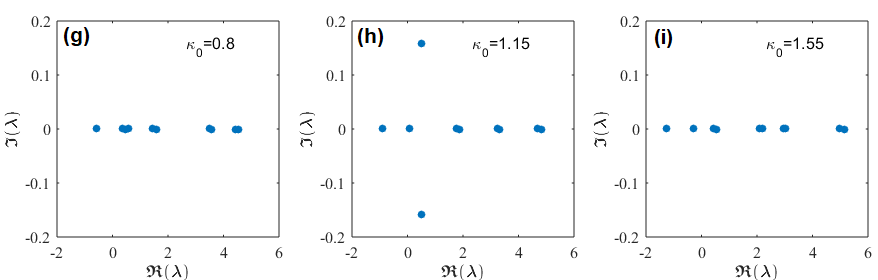}{0.9\linewidth}
\caption{Representative spectral scans under variation of a single control parameter. Top set: varying $\gamma$ at fixed $(W,\kappa_0)=(3,0.9)$. Middle set: varying $W$ at fixed $(\gamma,\kappa_0)=(0.2,0.9)$. Bottom set: varying $\kappa_0$ at fixed $(W,\gamma)=(2.2,0.25)$. In each case, the selected interval contains an unbroken--broken--unbroken $\PT$ sequence.}
\label{fig:varyParameters}
\end{figure*}

Figure~\ref{fig:varyParameters} gives direct spectral snapshots of three re-entrant sequences. In panels (a)--(c), the gain--loss strength is increased at fixed $(W,\kappa_0)=(3,0.9)$. At $\gamma=0.54$ all displayed eigenvalues lie on the real axis; at $\gamma=0.64$ one pair has moved to nonzero, opposite imaginary parts; and at $\gamma=0.74$ the pair has returned to the real axis. The corresponding sequence is therefore unbroken--broken--unbroken within this finite interval.

Panels (d)--(f) demonstrate that the same sequence can be driven geometrically. At fixed $(\gamma,\kappa_0)=(0.2,0.9)$, the spectra at $W=1.9$ and $W=2.7$ are real, while the intermediate value $W=2.3$ supports a complex-conjugate pair. One can visualize this re-entrant transition by looking at panel (c) of the Figure~\ref{fig:phaseMaps}. Here, we move along the horizontal line at a value of $\gamma = 0.2$.  We cross the boundaries of the unstable region, moving from one stable region to an unstable region, and then back to another stable one.

In our opinion, the most interesting case, from the experimental point of view is shown in panels (g)--(i) of the figure~\ref{fig:varyParameters}.  The image shows a coupling-amplitude scan at fixed $(W,\gamma)=(2.2,0.25)$. Again the spectra at $\kappa_0=0.8$ and $1.55$ are real, whereas $\kappa_0=1.15$ lies inside a broken-$\PT$ window. 

These examples show that re-entry is mode selective. Only the branch pair involved in the relevant exceptional-point contour becomes complex, while the remaining low-lying eigenvalues can stay real. The restored phase should therefore be understood as a bounded parameter window produced by a particular branch collision, not as stability for arbitrarily large gain--loss strength or coupling.

\section{Nonlinear continuation and propagation stability}
\label{sec:nonlinear}

\subsection{Stationary problem and continuation protocol}
\label{sec:continuation}

For $\sigma\neq0$, stationary states can be written as
\begin{equation}
\psi_j(x,t)=\phi_j(x)e^{-\ii\mu t},
\label{eq:stationaryAnsatz}
\end{equation}
where a bounded $\PT$-symmetric stationary state has real $\mu$. Substitution into Eqs.~\eqref{eq:model1}--\eqref{eq:model2} gives
\begin{align}
\mu\phi_1&=-\phi_1''+\ii\gamma\phi_1
+\frac{\sigma|\phi_1|^2}{1+\beta|\phi_1|^2}\phi_1
+\kappa(x)\phi_2,
\label{eq:stationary1}\\
\mu\phi_2&=-\phi_2''-\ii\gamma\phi_2
+\frac{\sigma|\phi_2|^2}{1+\beta|\phi_2|^2}\phi_2
+\kappa(x)\phi_1.
\label{eq:stationary2}
\end{align}
Direct numerical solution of the nonlinear eigenproblem defined by Eqs.~\eqref{eq:stationary1}--\eqref{eq:stationary2} is computationally demanding. We therefore employ a real-time adiabatic propagation method based on standard numerical propagation techniques for nonlinear Schrödinger equations \cite{Taha1984,Yang2010}, similar in spirit to that used for nonlinear coupled-ring dynamics in Ref.~\cite{Hung2017}. This approach provides a practical means of generating dynamically
accessible nonlinear waveforms from their linear counterparts. Specifically, an eigenstate of the linear system is first selected in the unbroken-$\PT$ phase. The nonlinear coefficient $\sigma(t)$ is then increased monotonically from zero to a target value $\sigma_f$ over a ramp time $t_{\mathrm{ramp}}$. After the ramp, $\sigma$ is held fixed and the propagation continues until $t_{\mathrm{end}}$. A state is regarded as propagation-stable over the simulated time interval if the powers in both rings remain bounded, no secular growth develops after the ramp, and the intensity profile remains stationary apart from an overall phase evolution. 
Although this approach does not enumerate all stationary solutions of Eqs.~\eqref{eq:stationary1}--\eqref{eq:stationary2} or replace a complete linear stability analysis, it identifies nonlinear branches that are dynamically accessible under physically realistic excitation conditions and remain propagation-stable over the simulated time interval. Accordingly, the limiting values reported below should be interpreted as the largest propagation-stable values obtained under the adopted ramping and observation protocol.

\subsection{Kerr nonlinear states}
\label{sec:kerr}

\begin{figure}[!t]
\centering
\safeimage{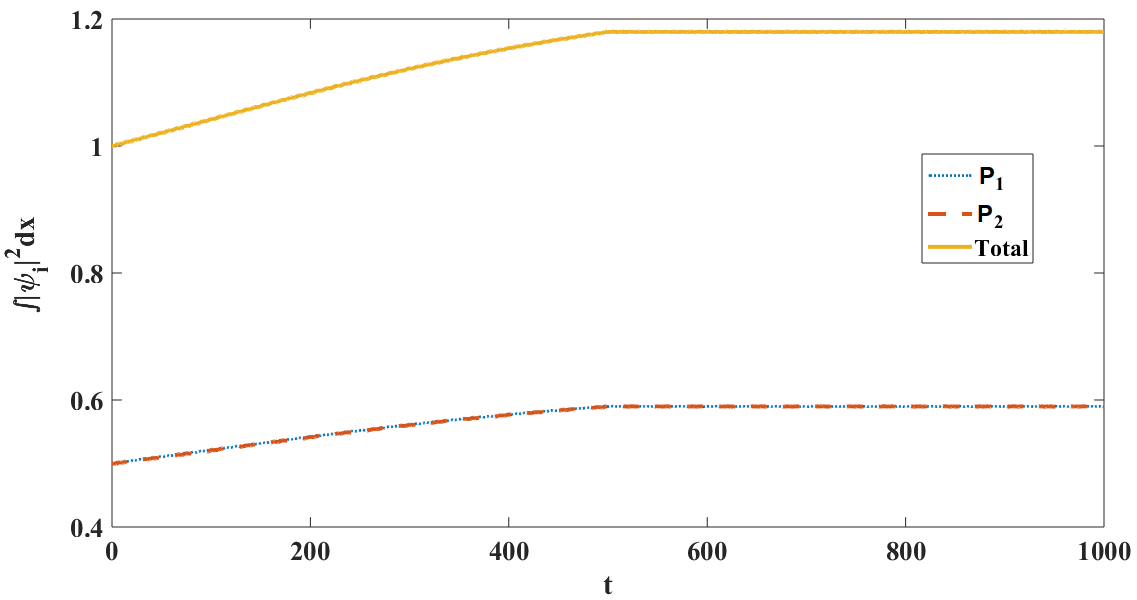}{1\columnwidth}
\safeimage{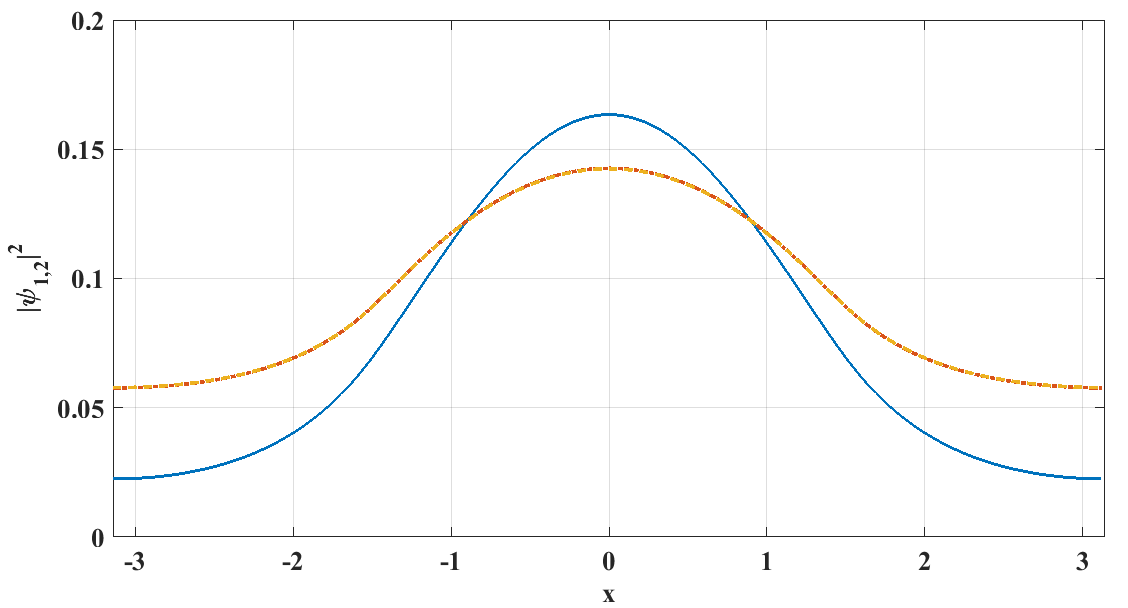}{0.98\columnwidth}
\caption{Adiabatic generation of a Kerr state from the linear ground mode. Upper panel: individual ring powers and their sum during the ramp and hold stages; the two individual powers nearly overlap and all curves settle after the ramp. Lower panel: linear input profile (solid) and final nonlinear profile (dashed). Parameters are $(W,\kappa_0,\gamma)=(1.5,0.9,0.3)$, $(\sigma_f,\beta)=(5,0)$, $t_{\mathrm{ramp}}=500$, and $t_{\mathrm{end}}=1000$. The largest propagation-stable value found with this protocol is approximately $\sigma_{f,\max}^{(g)}\simeq6$.}
\label{fig:KerrGround}
\end{figure}

We first consider the Kerr nonlinearity by setting $\beta=0$. As an illustration of the above numerical procedure, we choose a representative point from the phase map in Fig.~\ref{fig:phaseMaps}. The linear seed is taken at $(W,\kappa_0,\gamma)=(1.5,0.9,0.3)$, inside an unbroken-$\PT$ region. Starting from the ground mode, $\sigma$ is ramped to $\sigma_f=5$ by $t_{\mathrm{ramp}}=500$ and then held fixed until $t_{\mathrm{end}}=1000$. In the upper panel of Fig.~\ref{fig:KerrGround}, the individual ring powers remain nearly indistinguishable on the plotted scale while the total power increases during the ramp and reaches a constant plateau after $\sigma$ is fixed. According to Eq.~\eqref{eq:powerbalance}, this change of total power is produced by a small transient imbalance $P_1-P_2$ during the readjustment; the plateau indicates that the balance is restored after the ramp. The lower panel shows that the final intensity profile is broader and has a reduced central peak relative to the linear seed, but it exhibits no subsequent secular deformation over the observation interval. Repeating the calculation for increasing $\sigma_f$ gives propagation-stable ground-mode continuations up to approximately $\sigma_{f,\max}^{(g)}\simeq6$.

\begin{figure}[!t]
\centering
\safeimage{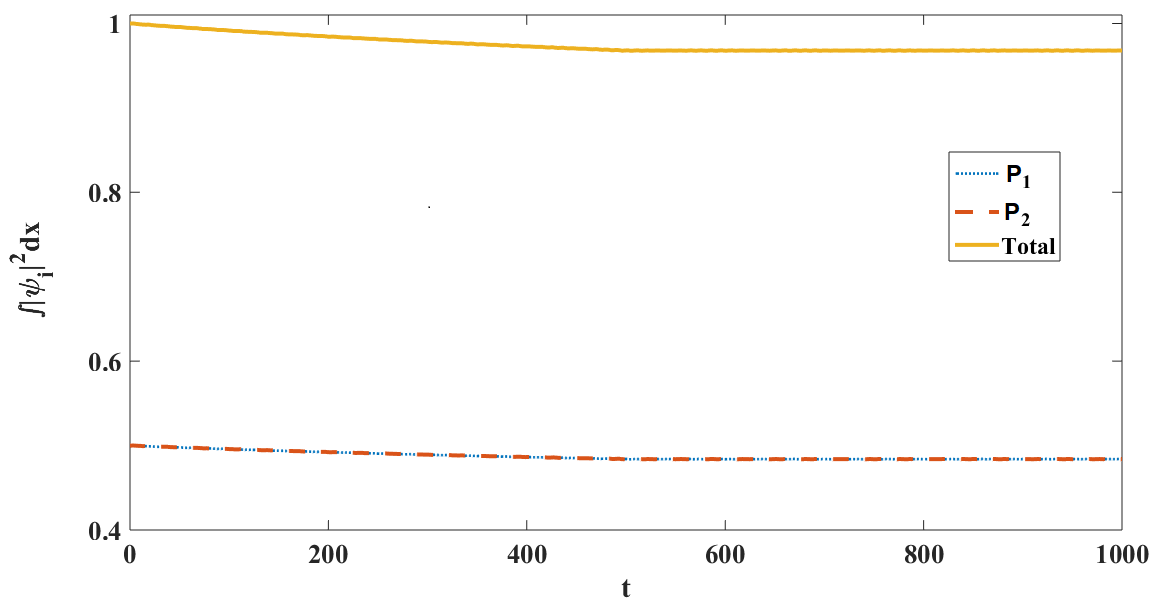}{1\columnwidth}
\safeimage{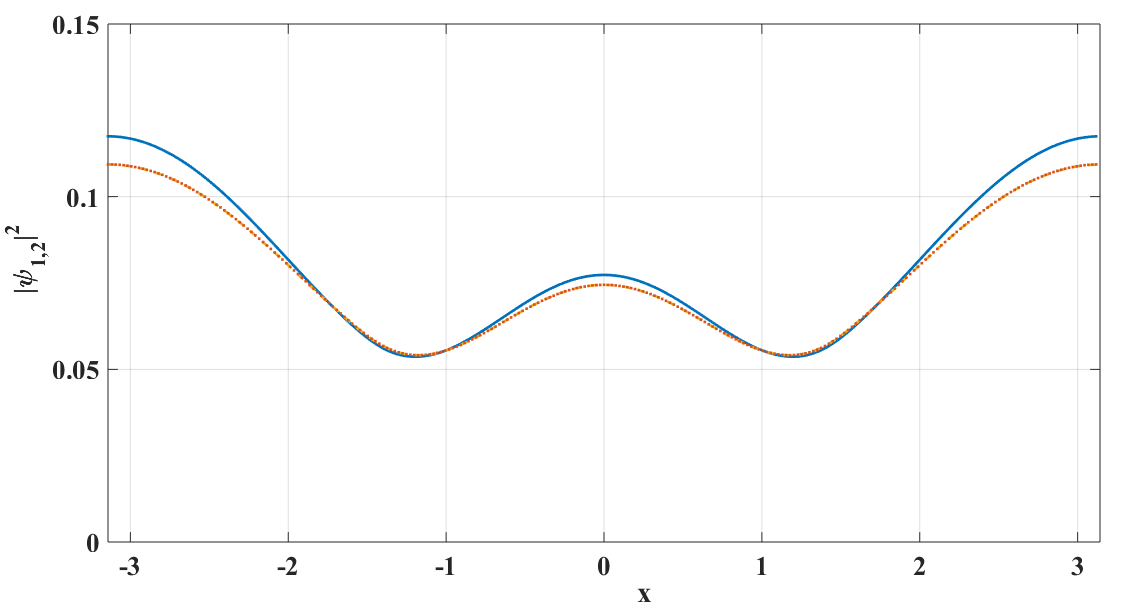}{0.969\columnwidth}
\caption{Adiabatic generation of a Kerr state from the first excited linear mode. Upper panel: individual and total powers, which approach a plateau after the ramp. Lower panel: linear input profile (solid) and final nonlinear profile (dashed); the multi-lobed excited-state structure is retained. Parameters are $(W,\kappa_0,\gamma)=(1.5,0.9,0.3)$ and $(\sigma_f,\beta)=(1.5,0)$. The largest propagation-stable value found with the same protocol is approximately $\sigma_{f,\max}^{(e)}\simeq2$.}
\label{fig:KerrExcited}
\end{figure}

The same protocol can be applied to excited modes. Starting from the first excited linear state and ramping to $\sigma_f=1.5$ produces the persistent waveform shown in Fig.~\ref{fig:KerrExcited}. The individual powers again remain nearly balanced, but the total power decreases during the ramp before reaching a steady plateau. The final profile preserves the characteristic multi-lobed structure of the excited seed, including the two off-center minima, while undergoing a smaller nonlinear reshaping than the ground mode. The largest propagation-stable value found for this branch is approximately $\sigma_{f,\max}^{(e)}\simeq2$.

The narrower interval is consistent with the greater sensitivity of an excited spatial structure to nonlinear detuning and coupling-induced mixing with nearby modes. It also shows that the nonlinear continuation range cannot be inferred from the linear $\PT$ threshold alone; it depends on the profile and spectral isolation of the seed branch.

\subsection{Saturable nonlinear states}
\label{sec:satu}
For comparison, we also carried out calculations using the same continuation procedure with a saturable nonlinearity ($\beta=1$) in place of the Kerr response, a nonlinear model that has also been investigated in balanced gain--loss couplers \cite{Abdullaev2020}. For both the ground-state and first-excited branches, the saturable model remained propagation-stable over a broader range of nonlinear strength than the Kerr model. For the representative linear point considered above, the largest propagation-stable nonlinear strengths were found to be approximately $\sigma_{f,\max}^{(g)}\simeq9$ for the ground-state branch and $\sigma_{f,\max}^{(e)}\simeq2.5$ for the first-excited branch. This behavior is consistent with the bounded nonlinear phase shift of the saturable response, which suppresses nonlinear frequency shifts at high intensities and thereby mitigates destabilizing nonlinear effects. Since these calculations are included only for qualitative comparison, no separate figures for the saturable model are presented.

\section{Conclusions}
\label{sec:conclusion}
We have investigated the linear and nonlinear behavior of two locally
coupled $\PT$-symmetric ring resonators with balanced gain and loss. The
inter-resonator interaction is represented by a high-order super-Gaussian
profile whose angular width provides a geometric way of interpolating
between two analytically tractable regimes. In the homogeneous-coupling
limit, each angular harmonic forms an independent $\PT$ dimer with the
threshold $\gamma=\kappa_0$. In the fixed-amplitude narrow-contact limit,
the integrated coupling vanishes, and the spectrum approaches the
uncoupled form $n^2\pm\ii\gamma$.

The finite-width regime displays spectral behavior that is absent in a
uniformly coupled ring dimer. The nonzero Fourier components of the local
coupling mix angular harmonics and lift the degeneracy of the
counterpropagating $\pm n$ modes. Reflection symmetry allows the resulting
states to be classified by parity, and each excited doublet is resolved
into four parity-dependent branches. Collisions among different branch
pairs generate multiple exceptional-point boundaries rather than a single
$\PT$ threshold. Consequently, the phase diagrams contain disconnected
broken-$\PT$ regions and exhibit re-entrant
unbroken--broken--unbroken transitions when the gain--loss strength,
coupling width, or peak coupling amplitude is varied. The continuous
recovery of the homogeneous- and narrow-contact spectra provides a
stringent consistency check on the finite-width numerical calculations.

We have also examined the nonlinear evolution of selected ground and
excited modes from the unbroken-$\PT$ regime. By gradually increasing the
nonlinear coefficient during real-time propagation, we obtained
finite-amplitude Kerr waveforms that remain dynamically persistent over
the simulated observation interval for finite ranges of nonlinear
strength. The accessible continuation range is strongly mode dependent
and is substantially narrower for the first-excited branch than for the
ground branch. Supporting calculations with a saturable response produce
broader ranges of dynamical persistence, consistent with the bounded
nonlinear phase shift of saturable media.

The nonlinear results demonstrate dynamical persistence over the simulated observation interval. However, these findings do not constitute a complete stability analysis \cite{Yang2010}. For locally coupled rings, such an analysis becomes numerically challenging.

Overall, the results demonstrate that engineering the spatial profile of
inter-resonator coupling provides an effective means of controlling
multimode spectral structure, exceptional-point boundaries, re-entrant
$\PT$ transitions, and nonlinear wave dynamics in non-Hermitian
ring-resonator systems.

\section*{Declaration of competing interest}
The authors declare that they have no known competing financial interests or personal relationships that could have appeared to influence the work reported in this paper.

\section*{Acknowledgment}
This research is funded by the Vietnam National Foundation for Science and Technology
Development (NAFOSTED) under grant number 103.01-2021.152.

\end{document}